\documentclass{PoS}

\usepackage{epsfig}

\def\as{\alpha_s}
\def\pb1{pb$^{-1}$}
\def\q2{Q^2}
\def\g2{GeV$^2$}
\def\etjb{E^{\rm jet}_{T,{\rm B}}}
\def\etalab{\eta^{\rm jet}_{\rm LAB}}
\def\kt{k_T}
\def\mj{M^{\rm jj}}
\def\cgh{\cos\gamma_h}
\def\etajb{\eta^{\rm jet}_{\rm B}}
\def\oass{{\cal O}(\as^2)}
\def\asz{\as(\mz)}
\def\asmzp#1#2#3#4#5{\asz = #1^{+#3}_{-#2}\ {\rm (exp.)}\ ^{+#5}_{-#4}\ {\rm (th.)}}
\def\mz{M_Z}
\def\eti{E^i_{T,{\rm B}}}
\def\etj{E^j_{T,{\rm B}}}
\def\oasss{{\cal O}(\as^3)}

\title{Jet cross sections in neutral current deep inelastic scattering
  at ZEUS and determination of $\as$}

\ShortTitle{Jet cross sections in NC DIS at ZEUS and $\as$}

\author{\speaker{Claudia Glasman}\\
        On behalf of the ZEUS Collaboration\\
        Universidad Aut\'onoma de Madrid, Spain\\
        E-mail: \email{claudia.glasman@desy.de}}

\abstract{The latest results on jet cross sections in neutral current
  deep inelastic $ep$ scattering from the ZEUS Collaboration are
  presented. The new results were used to perform stringent tests of
  perturbative QCD and extract precise values of the strong
  coupling. Also, the measurements have the potential to constrain further
  the parton distribution functions in the proton if included in QCD
  fits.
}

\FullConference{XVIII International Workshop on Deep-Inelastic
  Scattering and Related Subjects, DIS 2010\\ 
  April 19-23, 2010\\
  Firenze, Italy}

\begin{document}

{\bf Introduction.}
Jet production in neutral current (NC) deep inelastic scattering (DIS)
at order $\as$ in the Breit frame, in which the photon and the proton
collide head on, proceeds via the boson-gluon fusion and QCD Compton
processes. The jet production cross section can be written in
perturbative QCD (pQCD) as the convolution of the parton distribution
functions (PDFs) in the proton, determined from experiment, and the
partonic cross sections, calculable in pQCD.

QCD processes are dominant in hadron colliders and represent a
significant background to e.g. new physics searches at LHC.
Measurements of jet production in NC DIS at HERA provide a clean
hadron-induced reaction and a powerful tool to test pQCD calculations,
determine $\as$ and its energy evolution. In addition, these
measurements can constrain the proton PDFs, in particular the gluon
density, if incorporated, together with structure function data, in
the fits to extract the PDFs, as it has been done by the ZEUS
Collaboration. The result was a reduction of the gluon-density
uncertainty by up to a factor of two for mid- to high-$x$ values, a
region of phase space relevant for new physics searches at LHC.

The new measurements from the ZEUS experiment at HERA include
inclusive-jet and dijet cross sections with more than a three-fold
increase in statistics with respect to previous analyses; these data will
help to constrain further the proton PDFs. The measurements were also
used to perform precise tests of pQCD, extract $\as$ and test the
performance of new jet algorithms that have recently become available.

{\bf Constraints on the proton PDFs.}
Measurements of dijet cross sections~\cite{zeus-pub-10-005} were
performed using 374 \pb1\ of ZEUS data. The phase space of the
measurement is given by photon virtualities $125<\q2<20000$~\g2\ and
inelasticity $0.2<y<0.6$. The jets were searched using the $\kt$
cluster algorithm~\cite{np:b406:187} in the longitudinally invariant
inclusive mode~\cite{pr:d48:3160} and selected with $\etjb>8$ GeV and
$-1<\etalab<2.5$, where $\etjb$ is the jet transverse energy in the
Breit frame and $\etalab$ is the jet pseudorapidity in the laboratory
frame. A cut on the invariant mass of the dijet system, $\mj$, of 20
GeV was applied to remove the regions of phase space where the pQCD
calculations have limitations.

\begin{figure}
\setlength{\unitlength}{1.0cm}
\begin{picture} (18.0,5.5)
\put (1.5,-0.5){\epsfig{figure=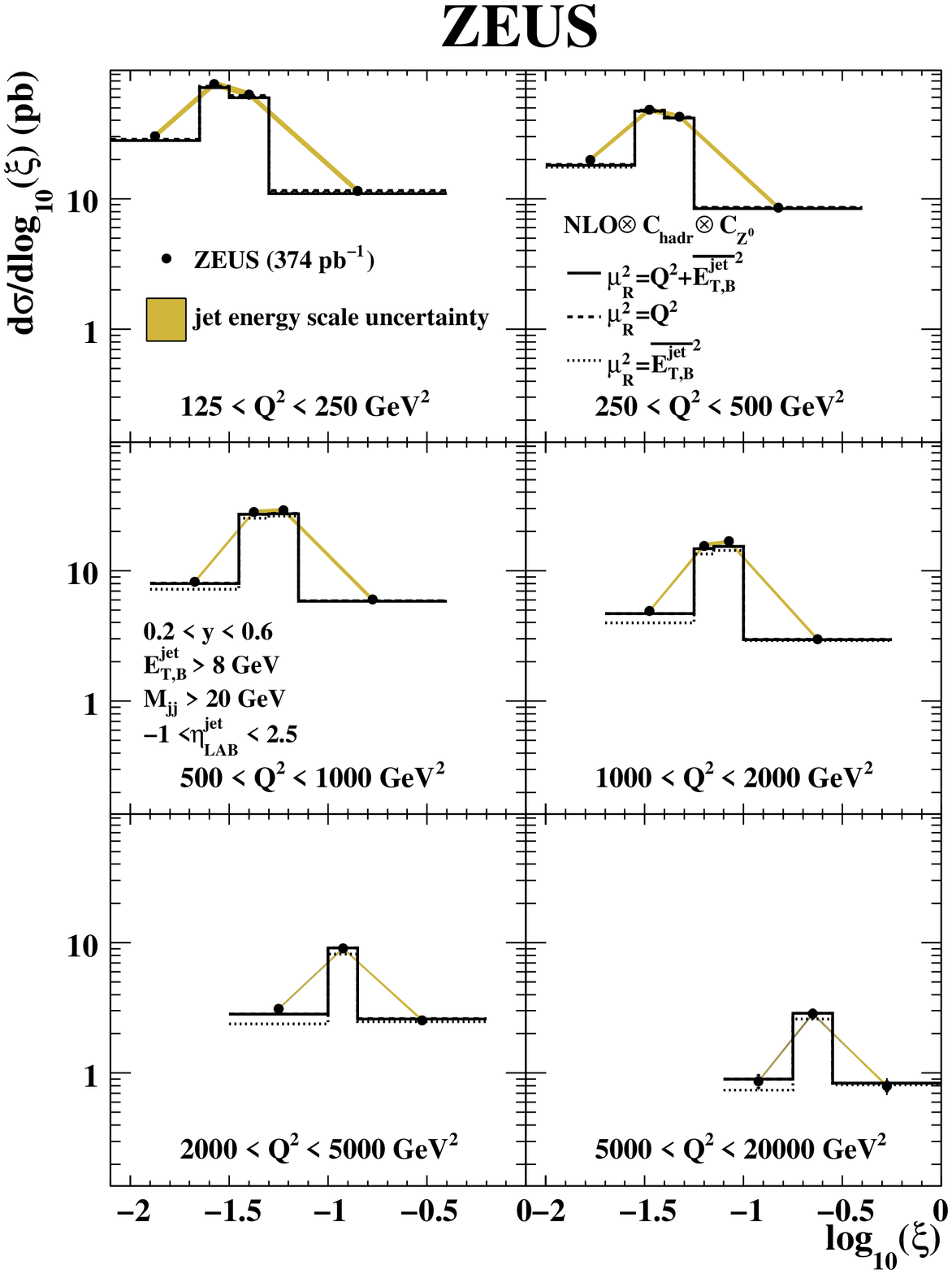,width=5cm}}
\put (7.5,-0.5){\epsfig{figure=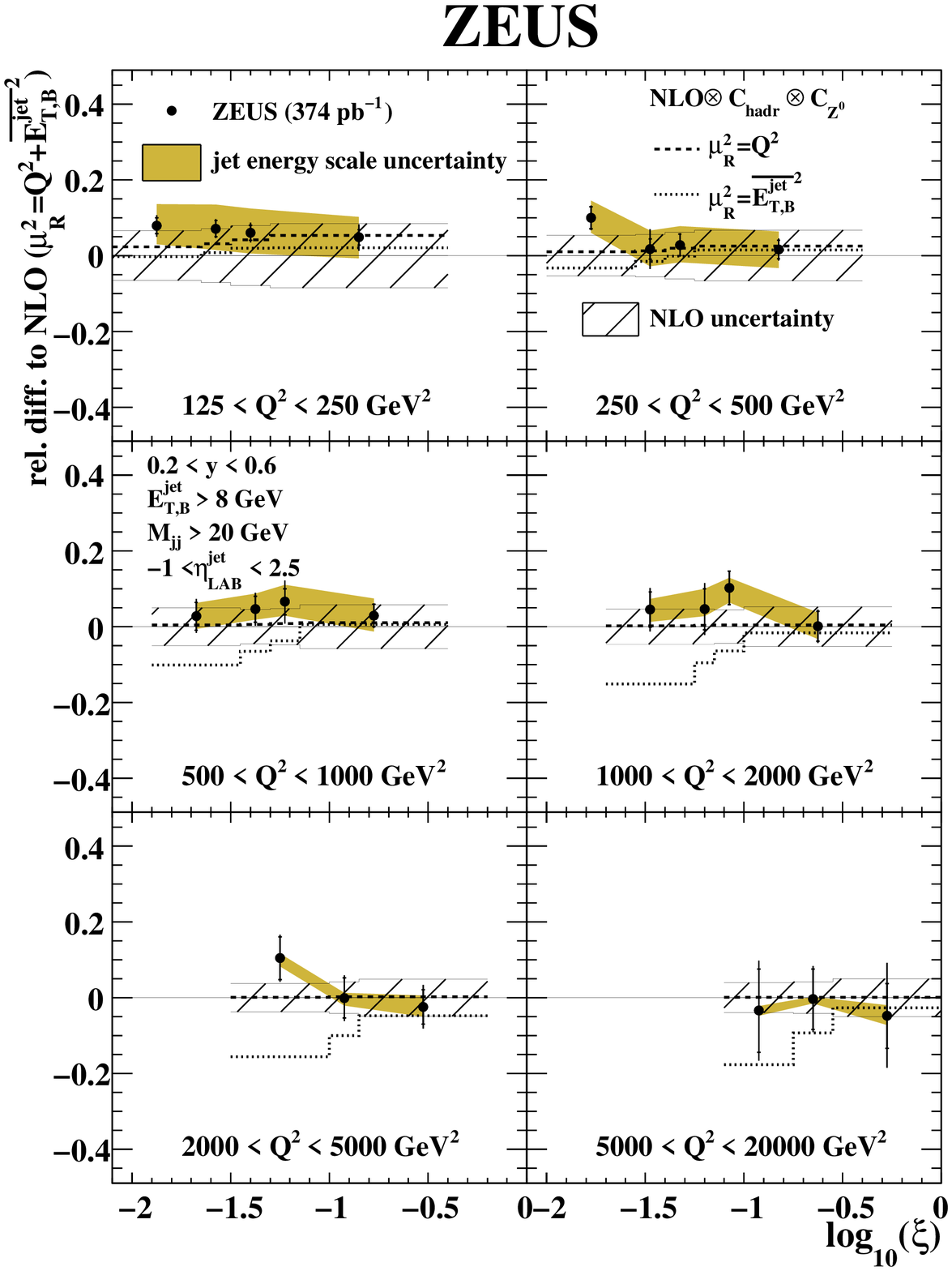,width=5cm}}
\end{picture}
\caption{\label{fig1}
{Dijet cross sections as functions of $\xi$ for different regions of $\q2$.
}}
\end{figure}

\begin{figure}
\setlength{\unitlength}{1.0cm}
\begin{picture} (18.0,5.5)
\put (1.5,-0.5){\epsfig{figure=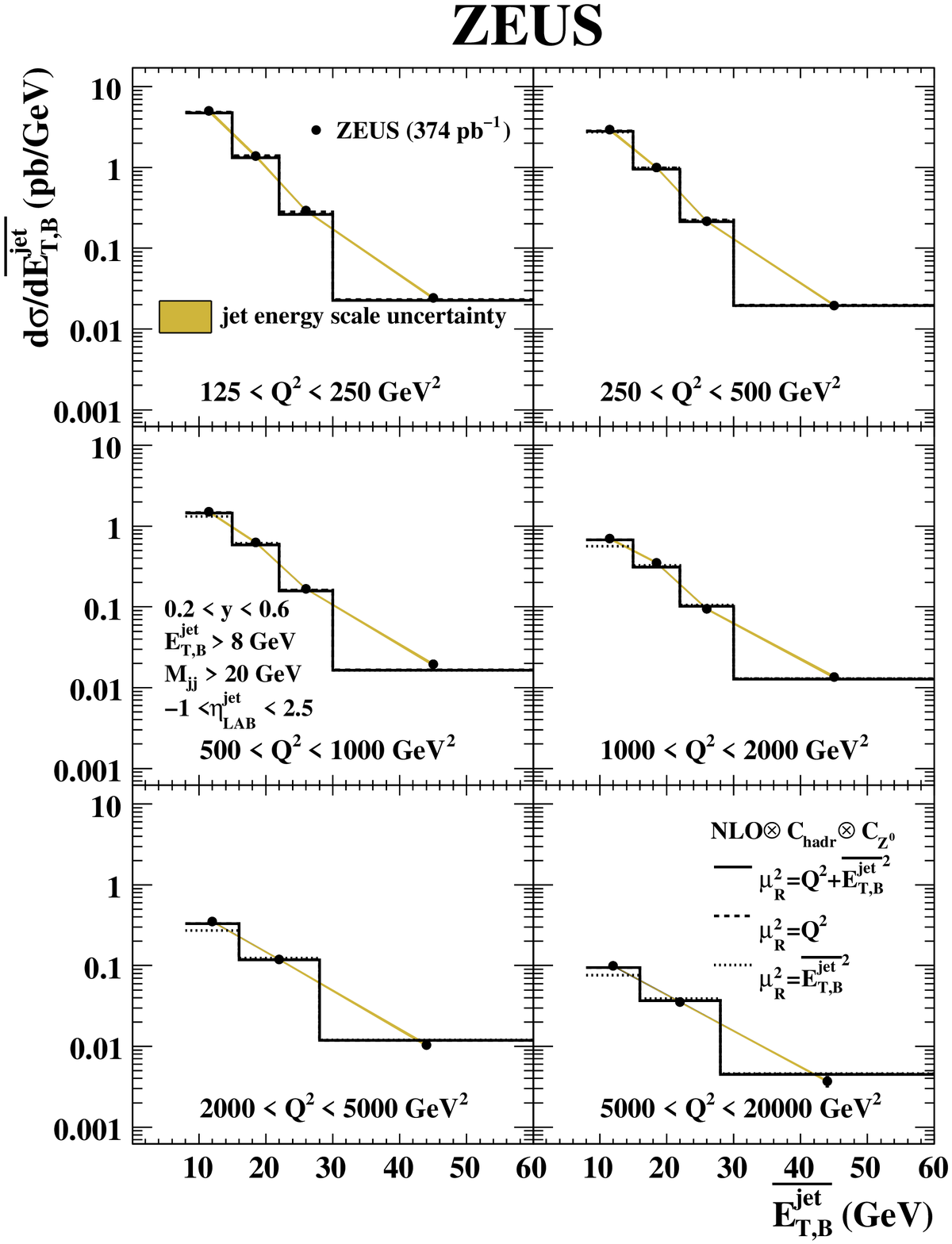,width=5cm}}
\put (7.5,-0.5){\epsfig{figure=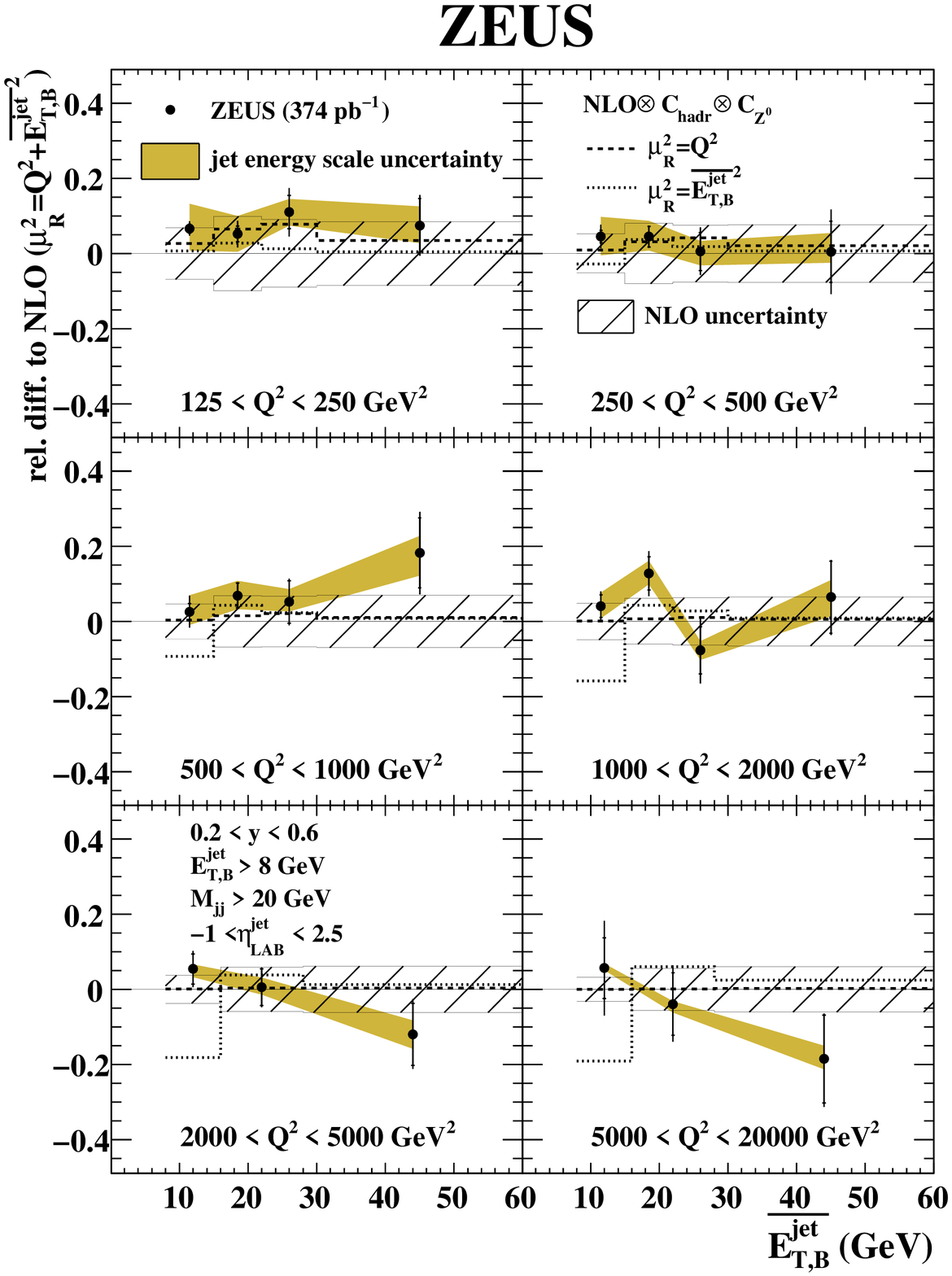,width=5cm}}
\end{picture}
\caption{\label{fig2}
{Dijet cross sections as functions of $\overline{\etjb}$ for different
  regions of $\q2$.
}}
\end{figure}

Figures~\ref{fig1} and \ref{fig2} show the dijet cross sections as
functions of $\xi=x_{\rm Bj}(1+(\mj)^2/\q2)$ and $\overline{\etjb}$,
the mean transverse energy of the two jets, in different regions of $\q2$,
respectively. The $\xi$ observable is an estimator of the fractional
momentum carried by the struck parton. The cross section as a function
of $\overline{\etjb}$ is well suited to make precise tests of pQCD.
The measured cross sections are very precise: the uncorrelated
uncertainties amount to $\sim 2\%$ at low $\q2$ and $\sim 8-10\%$ at
high $\q2$; the jet energy scale uncertainty, which has been reduced
to $\pm 1\%$, gives a contribution of $\pm 5\ (2)\%$ at low (high) $\q2$.
Next-to-leading-order (NLO) QCD predictions were computed using the
program {\sc Nlojet}++~\cite{prl:87:082001} with renormalisation
scale $\mu_R=\q2+(\overline{\etjb})^2$, factorisation scale $\mu_F=Q$
and the proton PDFs were parametrised using the CTEQ6.6~\cite{pr:d78:013004}
sets. The predictions give a good description of the data. To
ascertain the potential of the cross sections to constrain the gluon
density, the predicted gluon fraction and theoretical uncertainties
were studied in the phase-space region of the measurements: the
predicted gluon fraction is $\sim 75\%$ at low $\q2$ and decreases to 
$\sim 60\%$ for $\q2\sim 500$~\g2. The theoretical uncertainty
due to higher orders dominates in most of the phase-space region;
however, the PDF uncertainty is large in regions of phase space where
the gluon fraction is still sizeable and thus the high precision dijet
data presented have the potential to constrain further the proton PDFs.
Similar studies were performed for inclusive-jet cross sections as
functions of the $\etjb$ in different $\q2$
regions. Also in this case the PDF uncertainty is large in regions of 
phase space where the gluon fraction is still sizeable.

\begin{figure}
\setlength{\unitlength}{1.0cm}
\begin{picture} (18.0,5.0)
\put (1.5,-0.5){\epsfig{figure=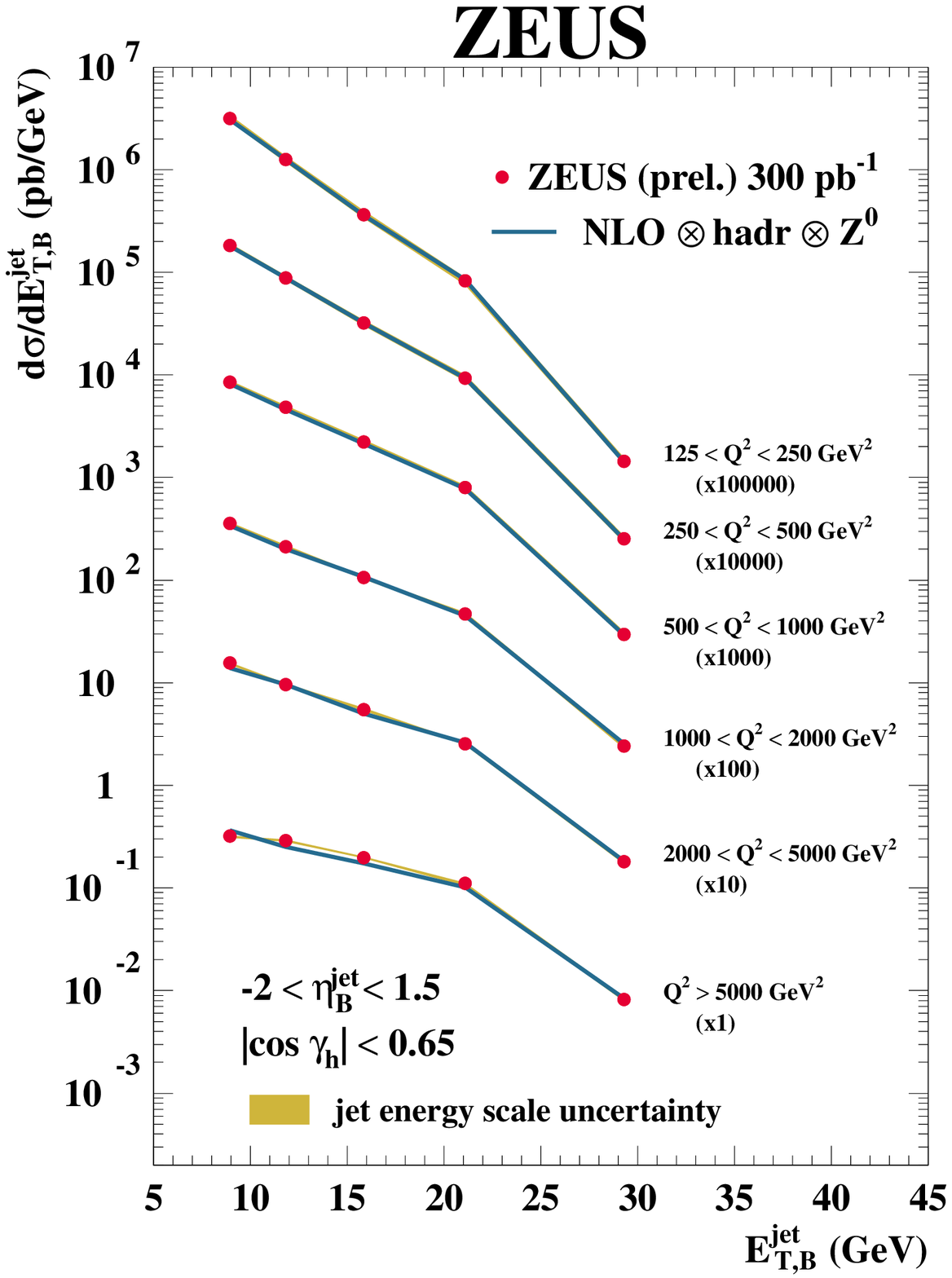,width=6cm}}
\put (7.5,-0.5){\epsfig{figure=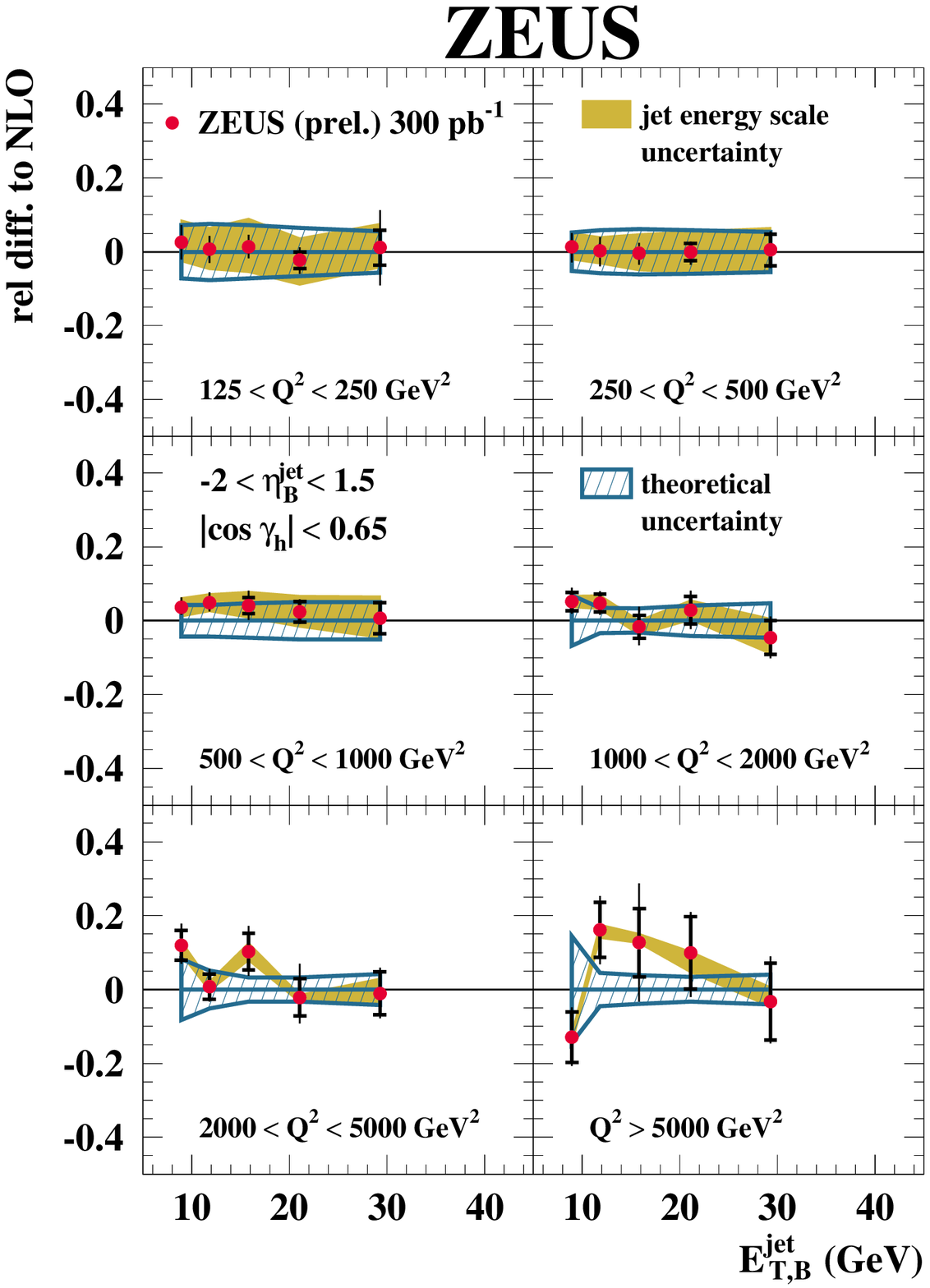,width=6cm}}
\end{picture}
\caption{\label{fig3}
{Inclusive-jet cross sections as functions of $\etjb$ for different
  regions of $\q2$.
}}
\end{figure}

Inclusive-jet cross sections were measured~\cite{zeus-prel-10-002}
using 300~\pb1\ of ZEUS data in the kinematic region of $\q2>125$~\g2\
and $\cgh<0.65$. Jets were searched in the Breit frame and selected
with $\etjb>8$ GeV and $-2<\etajb<1.5$. Figure~\ref{fig3} shows the
cross sections as functions of $\etjb$ in different regions of Q2. The
measured cross sections show that the $\etjb$ spectrum becomes harder
as $\q2$ increases. These data have also small experimental
uncertainties. NLO QCD calculations were computed using the program
{\sc Disent}~\cite{np:b485:291} with $\mu_R=\etjb$, $\mu_F=Q$ and the
ZEUS-S~\cite{pr:d67:012007} parametrisations of the proton PDFs. The
calculations describe the data very well in the whole measured range.
These measurements also have the potential to constrain further the
proton PDFs.

\begin{figure}
\setlength{\unitlength}{1.0cm}
\begin{picture} (18.0,5.5)
\put (-0.5,-0.2){\epsfig{figure=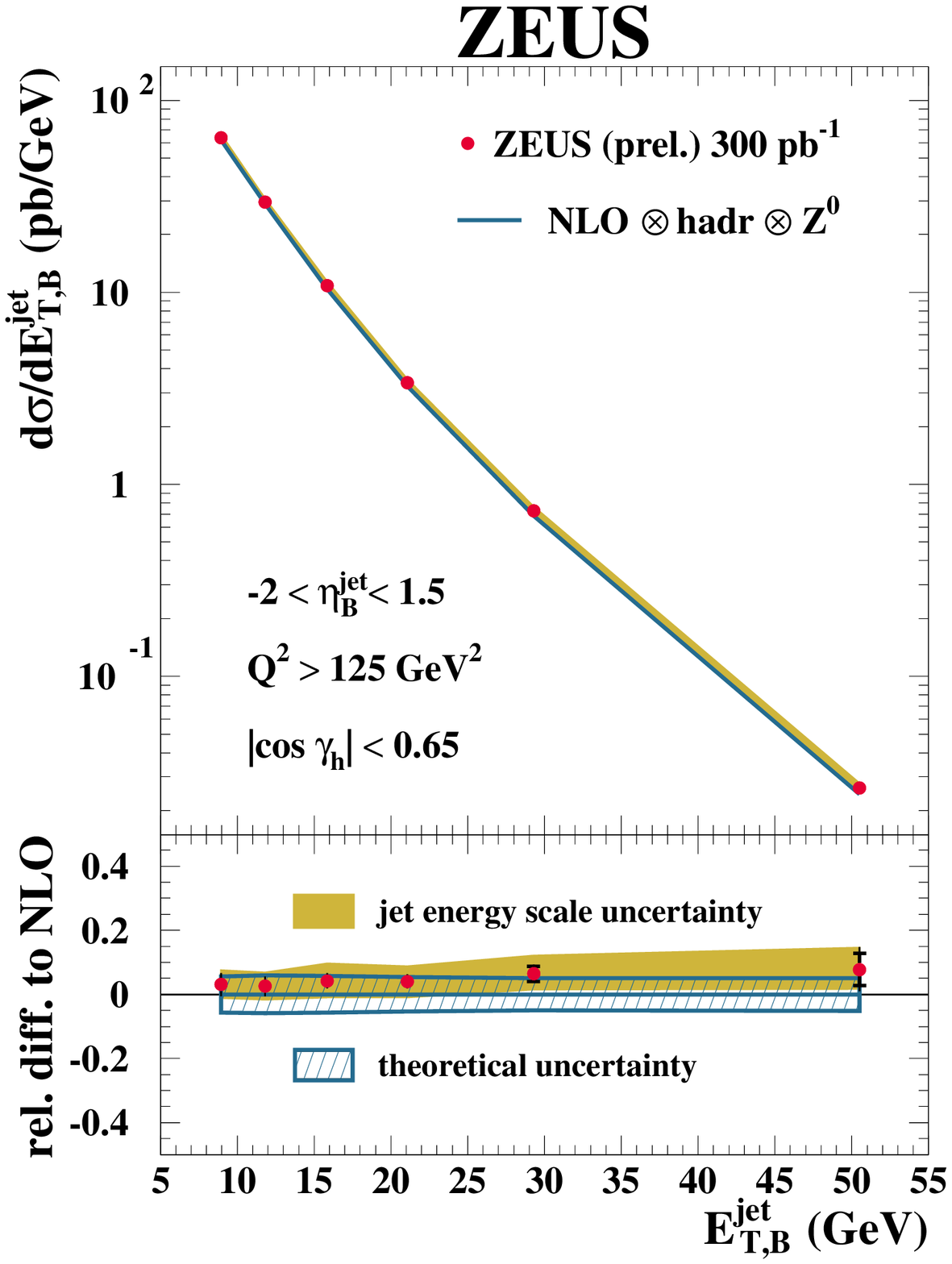,width=6cm}}
\put (4.5,-0.2){\epsfig{figure=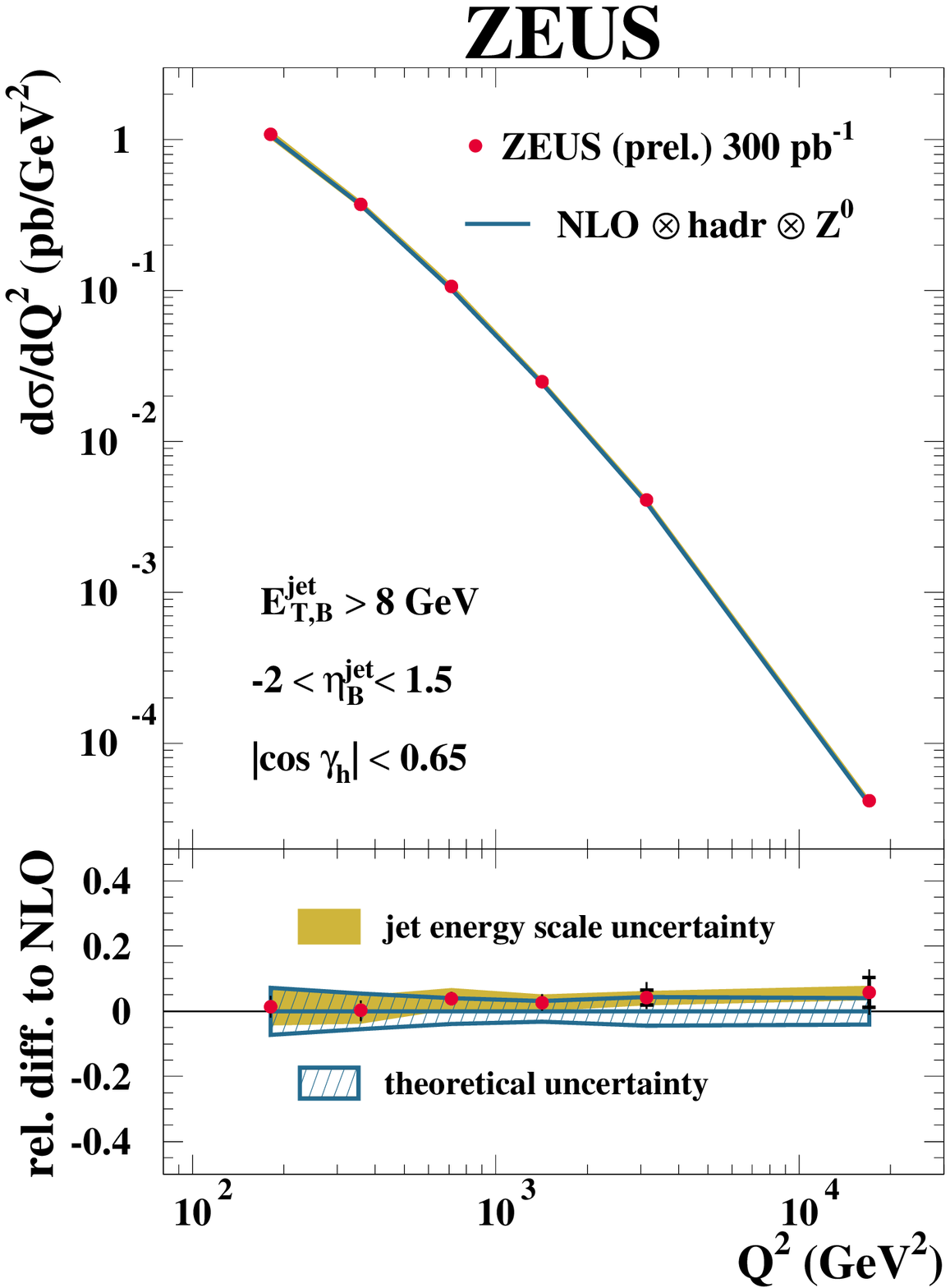,width=6cm}}
\put (9.5,-0.2){\epsfig{figure=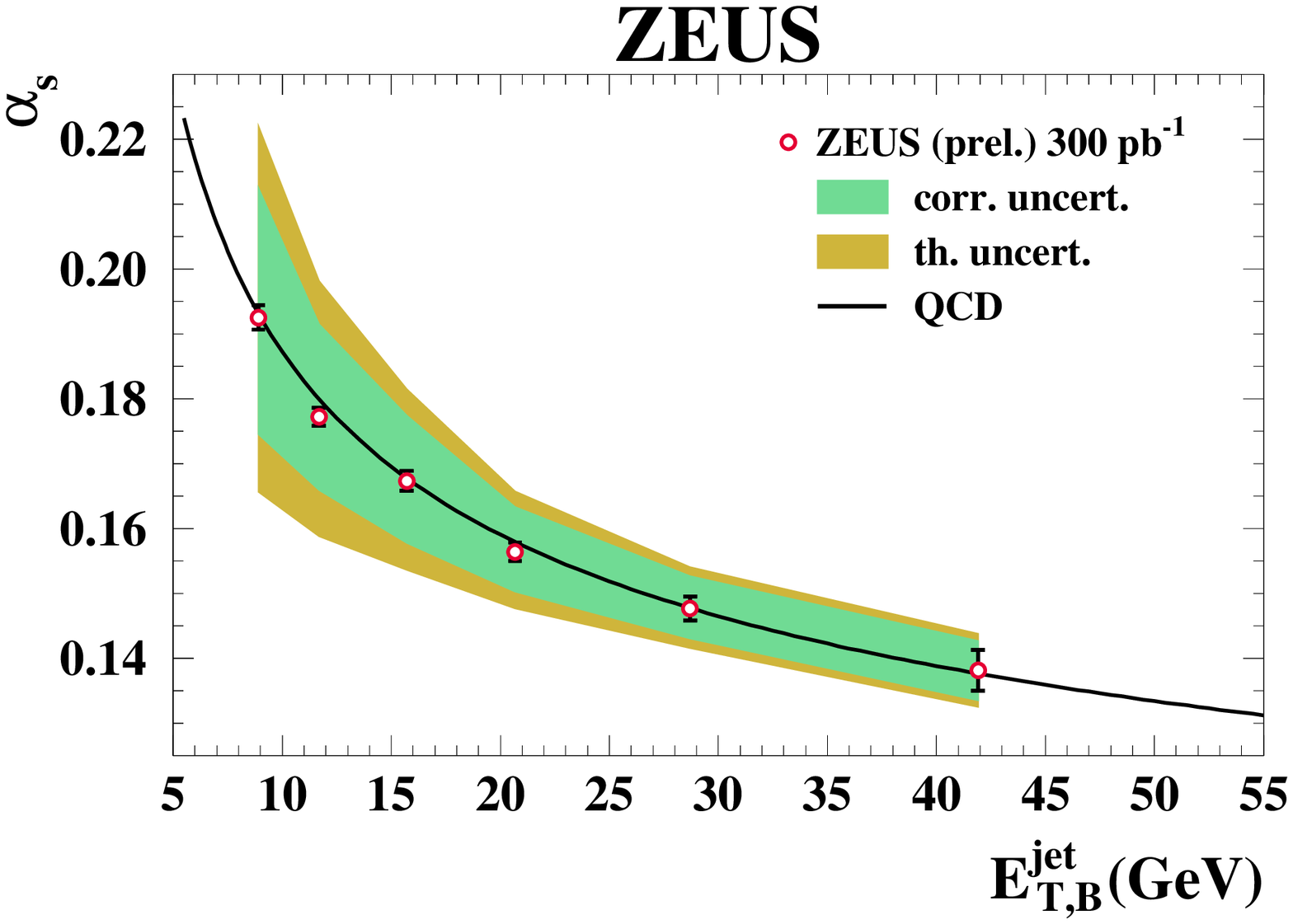,width=6cm}}
\put (2.7,-0.3){\small (a)}
\put (7.7,-0.3){\small (b)}
\put (12.7,-0.3){\small (c)}
\end{picture}
\caption{\label{fig4}
{Inclusive-jet cross sections as functions of (a) $\etjb$ and (b)
  $\q2$. (c) Energy-scale dependence of $\as$.
}}
\end{figure}

{\bf Tests of pQCD.}
Single-differential inclusive-jet cross sections were
measured~\cite{zeus-prel-10-002} as functions of $\etjb$ and $\q2$ to
perform stringent tests of pQCD. The advantages of using inclusive-jet
cross sections for performing such tests come from the fact that they
are infrared insensitive (no asymmetric cuts on $\etjb$ or mass cuts
are needed) and so a wider phase space is accessible than for dijet
cross sections and they present smaller theoretical
uncertainties. Also, these cross sections are suited to test resummed
calculations. Figures~\ref{fig4}a and \ref{fig4}b show the cross
sections as functions of $\etjb$ and $\q2$, respectively. The measured
cross section decreases by more than three (five) orders of magnitude
within the measured range and have small experimental
uncertainties. The theoretical uncertainties are also small and
dominated by the terms beyond NLO; this uncertainty is smaller than
$5\%$ for $\q2>250$~\g2. The NLO calculations describe very well both
measured distributions. This demonstrates the validity of the
description of the dynamics of inclusive-jet production by pQCD at order
$\oass$. These cross sections are directly sensitive to $\as$ and
present small experimental and theoretical uncertainties, therefore
they are particularly suited to determine this fundamental parameter.

A value of $\asz$ was determined from a NLO QCD fit to the data for
$\q2>500$~\g2: $\asmzp{0.1208}{0.0032}{0.0037}{0.0022}{0.0022}$.
In the fitting procedure, the running of $\as$ as predicted by QCD was
assumed. The experimental uncertainties are dominated by the jet
energy scale and amounts to $\pm 1.9\%$. The theoretical uncertainties
are dominated by the terms beyond NLO and amounts to $\pm
1.5\%$. Other contributions to the theoretical uncertainties are:
proton PDFs ($\pm 0.7\%$), hadronisation corrections ($\pm 0.8\%$) and
variation of $\mu_F$ (negligible). Therefore, a very precise value of
$\asz$ was obtained from the inclusive-jet cross sections in NC DIS
with a total uncertainty of $\sim 3.5\%$, with a contribution of only
$\sim 1.9\%$ from the theoretical uncertainties.
The energy-scale dependence of $\as$ was also determined from a NLO
QCD fit to these data. Values of $\as$ were extracted at each mean
value of $\etjb$ measured without assuming the running of $\as$. The
results are shown in Fig.~\ref{fig4}c together with the correlated
(inner band) and the theoretical (outer band) uncertainties. The
black curve represents the QCD prediction for the running of
$\as$. The $\etjb$-dependence of the extracted values of $\as$ is in
very good agreement with the predicted running of $\as$ over a large
range in $\etjb$.

Testing pQCD with jets requires infrared- and collinear-safe jet
algorithms. Up to now, only the $\kt$ algorithm fulfilled these
requirements at all orders. This algorithm has been tested extensively
at HERA and it was proven that it has a good performance with small
theoretical uncertainties and hadronisation corrections. Recently, new
infrared- and collinear-safe jet algorithms, namely the
anti-$\kt$~\cite{jhep:04:063} and SIScone~\cite{jhep:05:086}, have
been developed. 
Cluster algorithms, such as the $\kt$ and anti-$\kt$ jet algorithms,
combine particles according to their distance in the $\eta-\phi$ plane
via $d_{ij}={\rm min}((\eti)^{2p},(\etj)^{2p})\cdot\Delta R^2/R^2$, in 
which the parameter $p$ is set to 1 for the $\kt$ and to $-1$ for the
anti-$\kt$. The anti-$\kt$ algorithm is also infrared and collinear
safe to all orders and, contrary to the $\kt$, provides approximately
circular jets, which is experimentally desirable to obtain stable
detector corrections. The SIScone algorithm is a seedless cone
algorithm and, contrary to other versions of cone algorithms, is
infrared and collinear safe to all orders.

\begin{figure}
\setlength{\unitlength}{1.0cm}
\begin{picture} (18.0,6.0)
\put (0.0,-0.5){\epsfig{figure=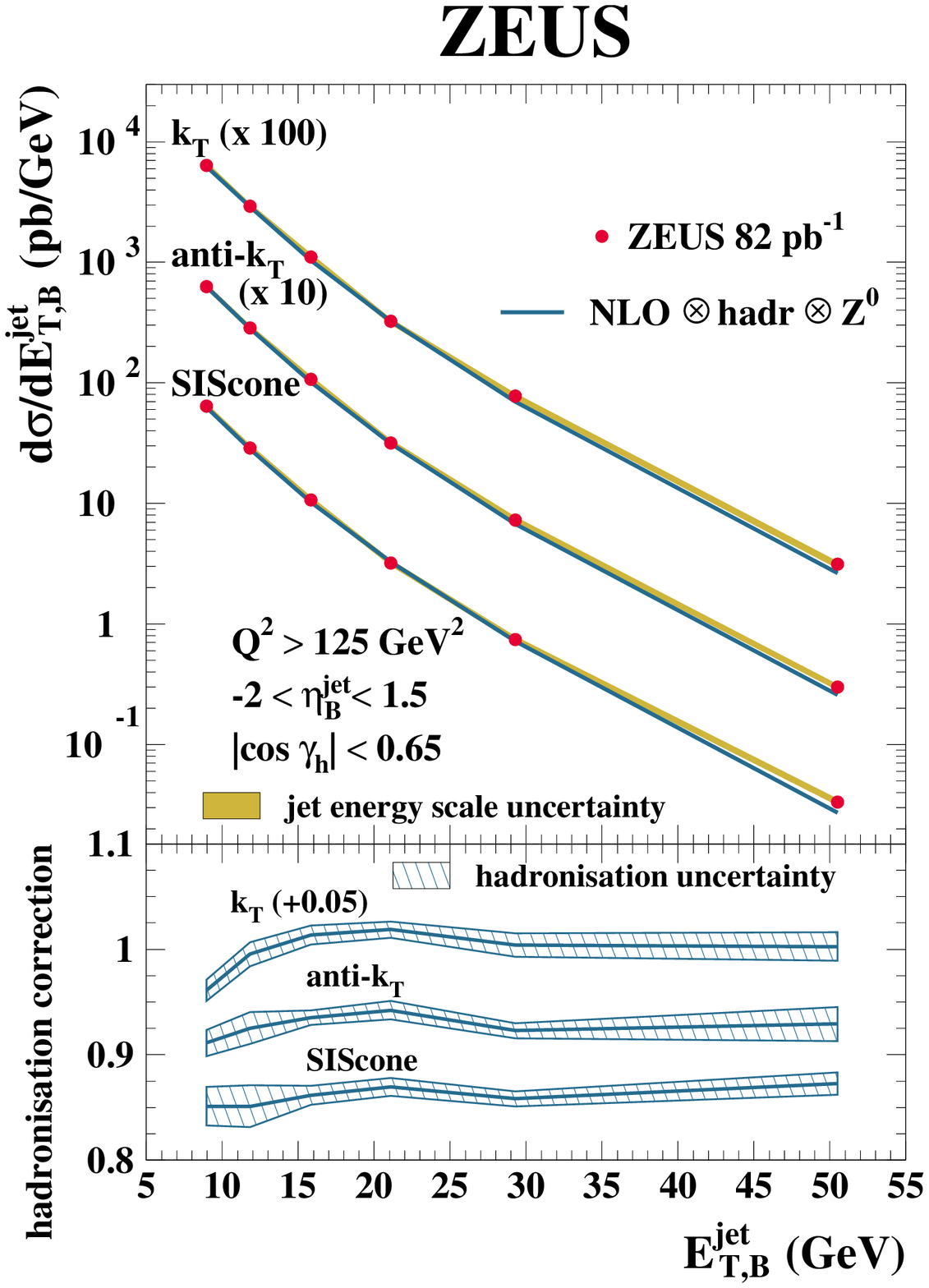,width=6cm}}
\put (5.0,-0.5){\epsfig{figure=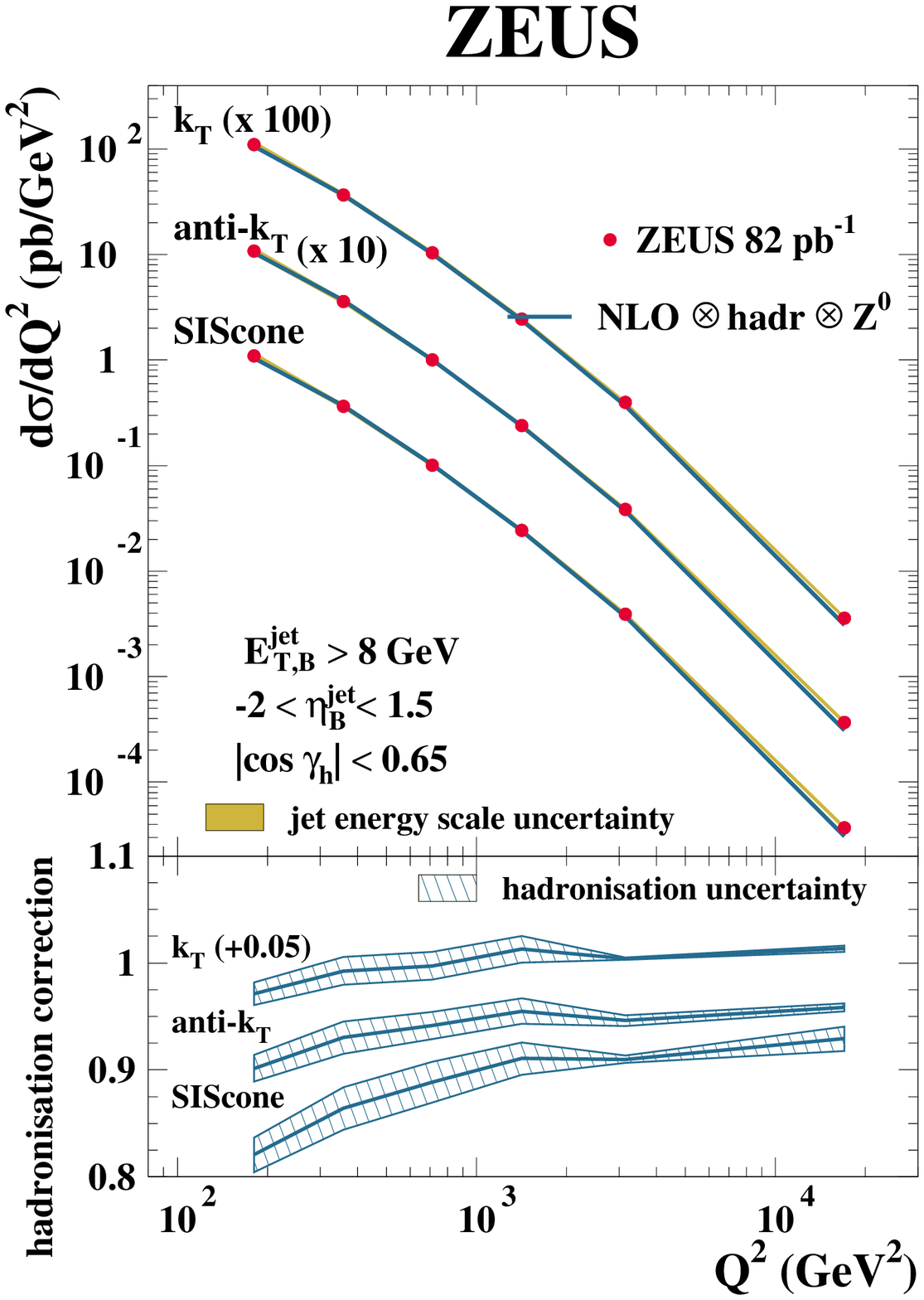,width=6cm}}
\put (9.5,0.5){\epsfig{figure=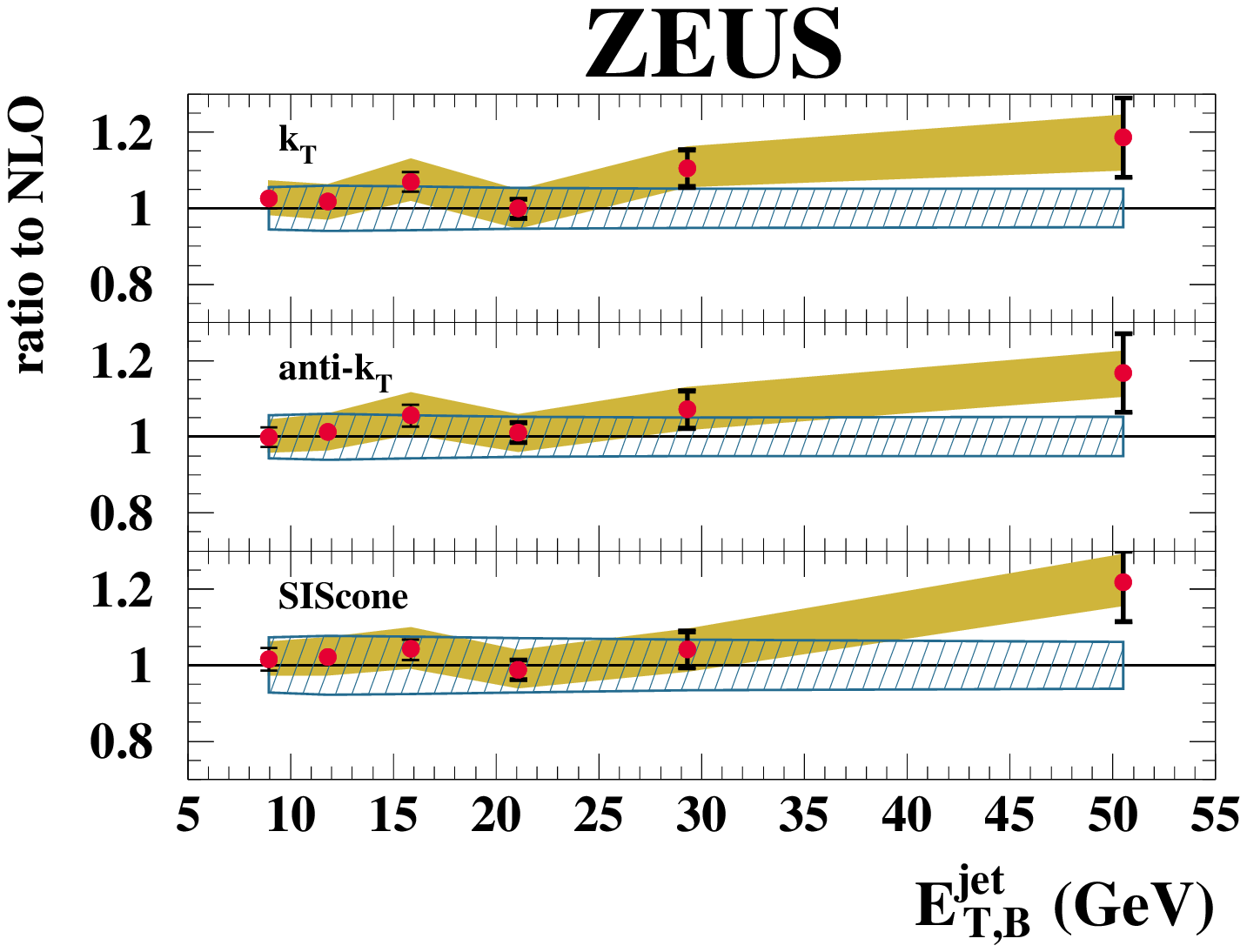,width=6cm}}
\put (9.5,-3.0){\epsfig{figure=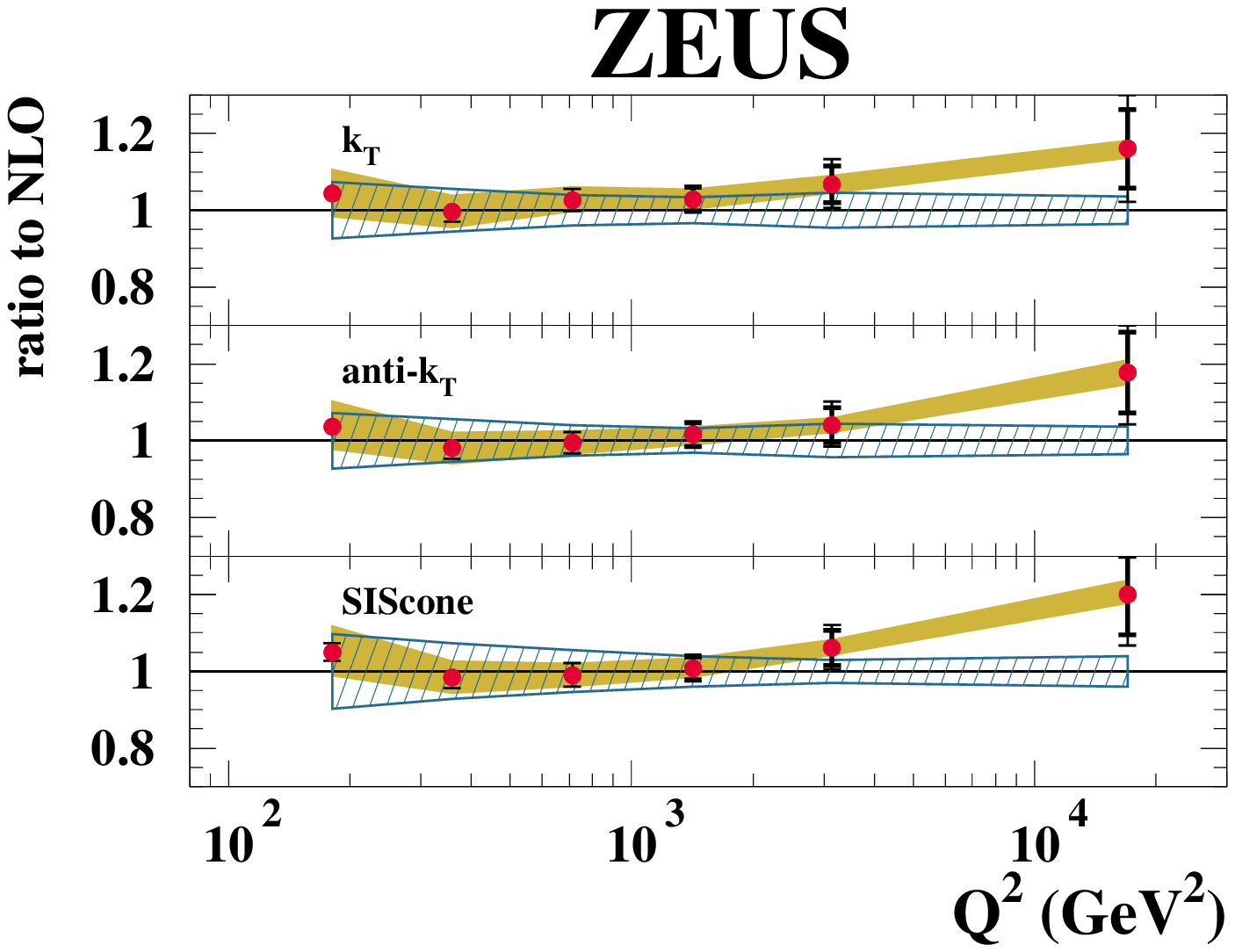,width=6cm}}
\end{picture}
\caption{\label{fig5}
{Inclusive-jet cross sections based on different jet algorithms.}}
\end{figure}

Studies~\cite{desy-10-034}
were performed with ZEUS data to validate these algorithms for their
use in more complicated environments, such as hadron-hadron
colliders. The performance of the anti-$\kt$ and SIScone algorithms
was tested in the well-understood hadron-induced NC DIS process by
comparing measurements based on the new algorithms with those based on
the $\kt$ and by comparing the data and the pQCD
predictions. 
The theoretical uncertainties for these new jet algorithms were
studied and compared with those for the $\kt$ algorithm in inclusive-jet
cross sections. The uncertainties from the proton PDFs and the value
of $\as$ are very similar for all three jet algorithms. The
uncertainty from the terms beyond NLO and the modelling of the parton
shower are very similar for the $\kt$ and anti-$\kt$, but slightly
larger for the SIScone algorithm.

The inclusive-jet cross sections were measured as functions of
$\etjb$ and $\q2$ using the three jet algorithms (see
Fig.~\ref{fig5}). The shape and normalisation of the measured and
predicted cross sections are very similar; the data are very well
described by the NLO calculations. The hadronisation-correction factor
applied to the calculations, also shown in Fig.~\ref{fig5}, are
similar for the $\kt$ and anti-$\kt$ and somewhat bigger for the
SIScone algorithm.

\begin{figure}
\setlength{\unitlength}{1.0cm}
\begin{picture} (18.0,4.0)
\put (0.5,-3.5){\epsfig{figure=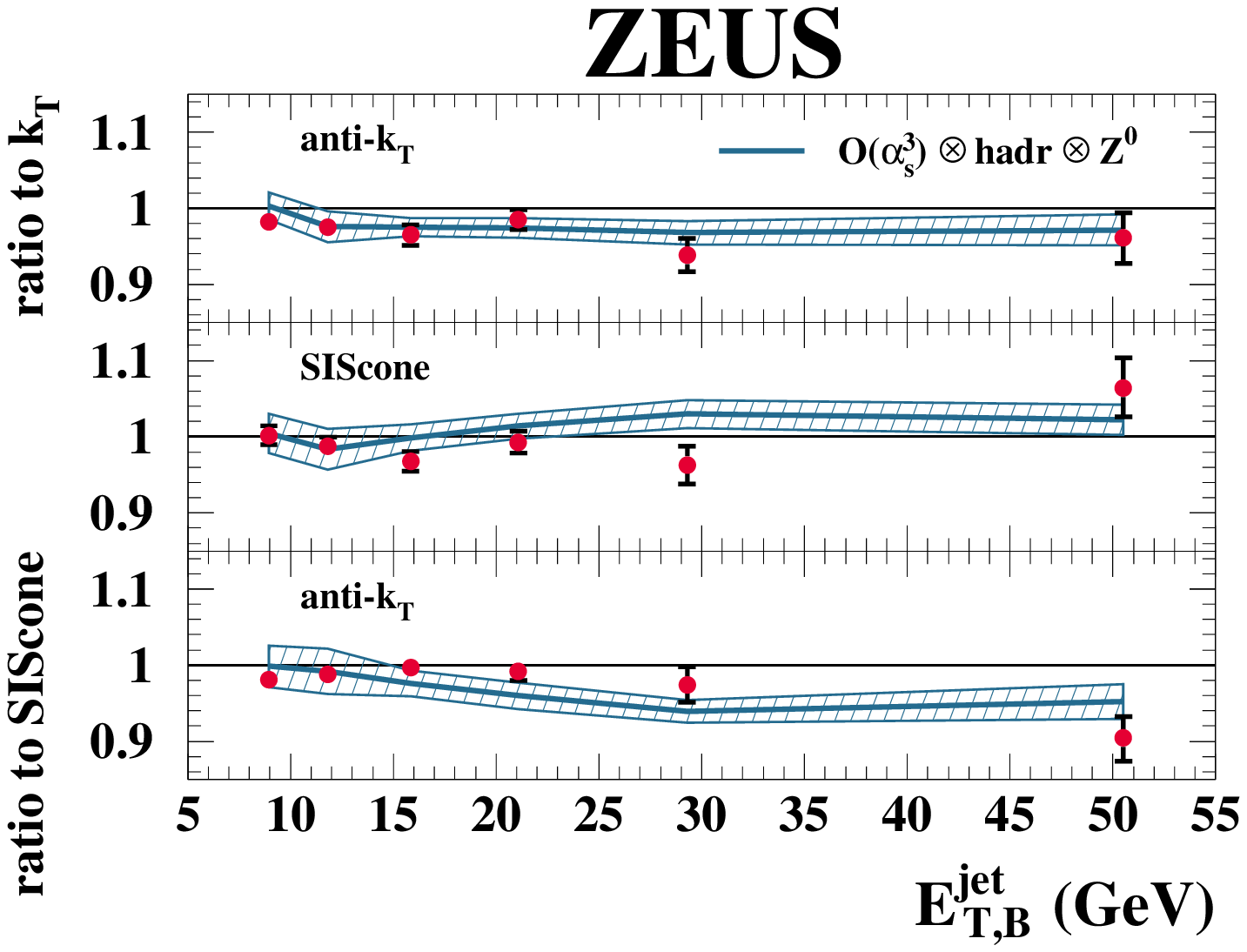,width=8cm}}
\put (6.5,-3.5){\epsfig{figure=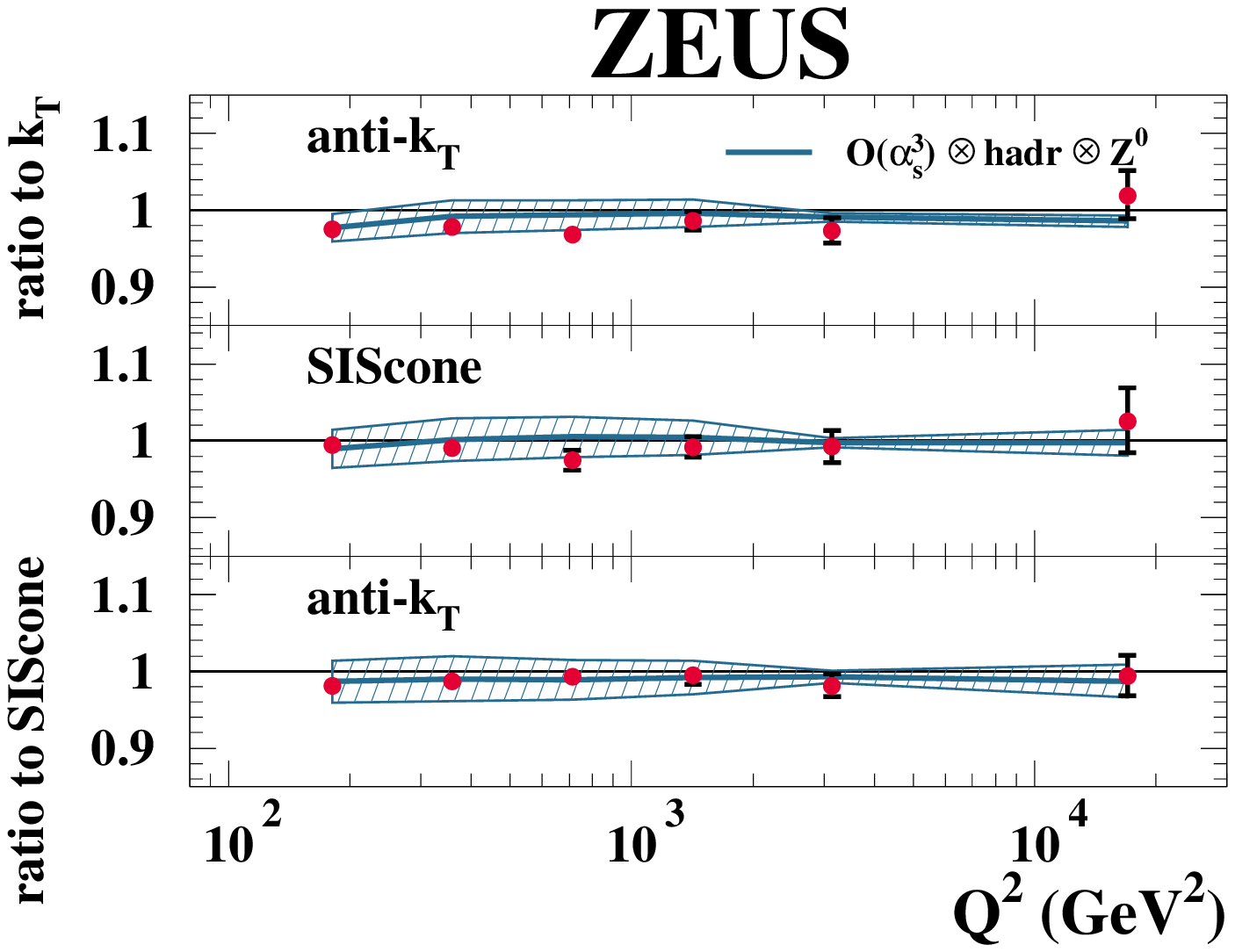,width=8cm}}
\end{picture}
\caption{\label{fig6}
{Ratios of cross sections based on different jet algorithms.}}
\end{figure}

To study in more detail the performance of the new algorithms,
the ratios of the cross sections between different algorithms were
measured. Inclusive-jet cross sections can be calculated only up to 
$\oass$ using the currently available programs. However, 
differences between cross sections using different algorithms can be
calculated up to $\oasss$ using {\sc Nlojet}++. In the case of the
SIScone, differences with the $\kt$ algorithm appear first for final
states with three partons, and in the case of the anti-$\kt$,
differences with the $\kt$ algorithm appear first for final states
with four partons.
Figure~\ref{fig6} shows the measured ratios for anti-$\kt$/$\kt$,
SIScone/$\kt$ and anti-$\kt$/SIScone as functions of $\etjb$ and $\q2$
together with the $\oasss$ predictions. The measured cross sections
show differences below $\sim 3.2\%$ as a function of $\q2$ and below
$3.6\%$ as a function of $\etjb$. The QCD predictions up to $\oasss$
give a good description of the measured ratios. The theoretical
uncertainty due to higher orders of the $\oasss$ calculation is
reduced and so the dominant uncertainty is that due to the QCD-cascade
modelling. These results demonstrate the ability of the pQCD
calculations including up to four partons in the final state to
account adequately for the differences between the jet algorithms.

Values of $\asz$ were extracted from the measured cross sections
using the three jet algorithms. The values obtained are: 
$\asmzp{0.1188}{0.0035}{0.0036}{0.0022}{0.0022}$ (anti-$\kt$),
$\asmzp{0.1186}{0.0035}{0.0037}{0.0026}{0.0026}$ (SIScone) and
$\asmzp{0.1207}{0.0036}{0.0038}{0.0023}{0.0022}$ ($\kt$).
These determinations are consistent with each other and have a similar
precision.

\begin{figure}
\setlength{\unitlength}{1.0cm}
\begin{picture} (18.0,6.0)
\put (-0.5,-0.5){\epsfig{figure=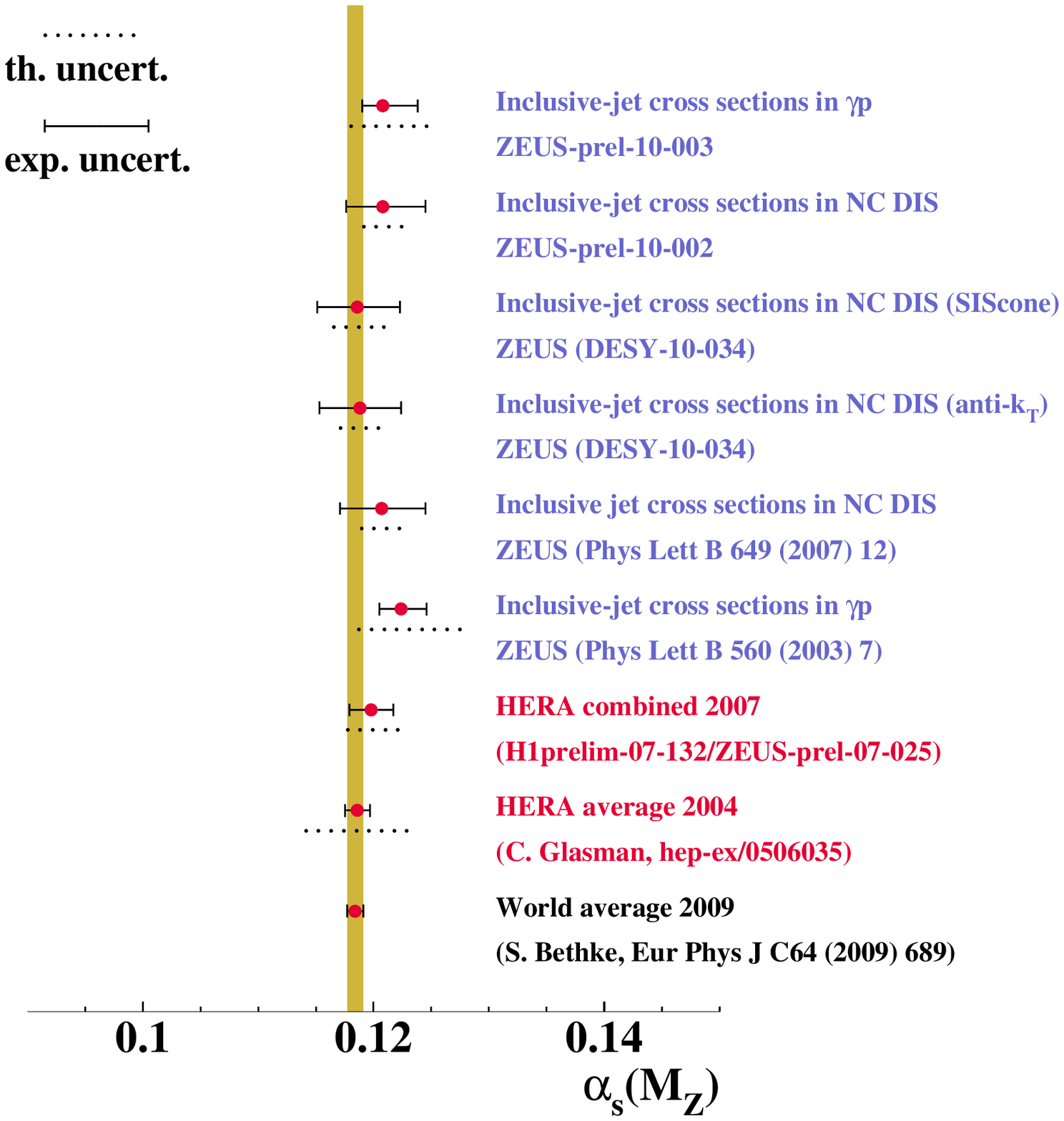,width=8cm}}
\put (7.5,-0.5){\epsfig{figure=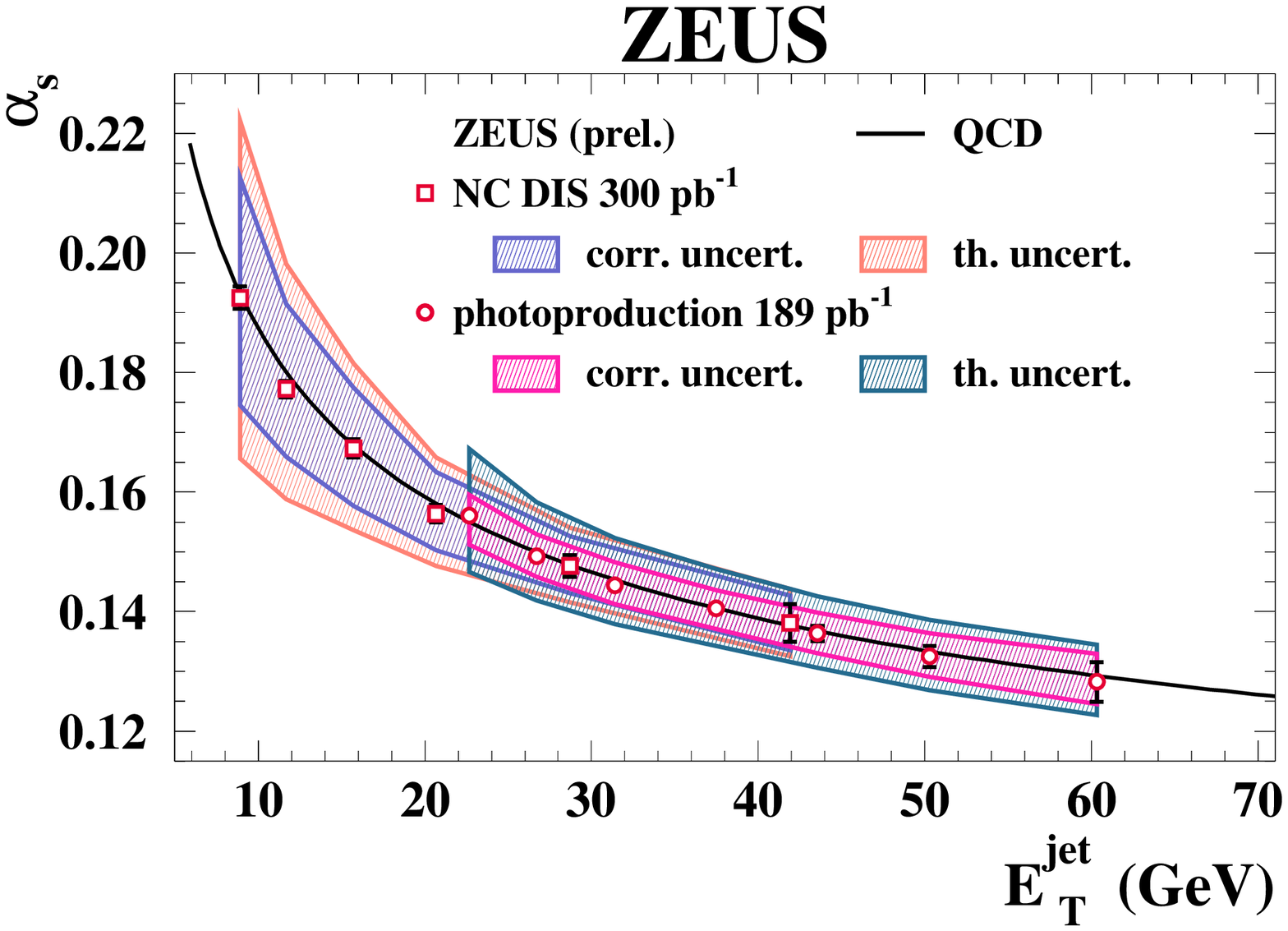,width=8cm}}
\put (2.4,-0.3){\small (a)}
\put (12.0,-0.3){\small (b)}
\end{picture}
\caption{\label{fig7}
{(a) Summary of $\asz$ values extracted from ZEUS data together with
  the HERA and world averages. (b) Summary of the running $\as$ values
extracted from ZEUS data.}}
\end{figure}

{\bf Summary.}
Figure~\ref{fig7}a shows a summary of the values of $\asz$ presented
together with other determinations from ZEUS, both in DIS and
photoproduction, and the HERA averages of 2004~\cite{hep-ex-0506035}
and 2007~\cite{proc:eps:2007:022013} and the current world
average~\cite{ppnp:58:351}. The measurements are consistent 
with each other and the world average. The summary of the
running of $\as$ from DIS data together with the results from
photoproduction is shown in Fig.~\ref{fig7}b. The measurements are
consistent with the predicted running of $\as$ over a wide range of
the scale. In addition, precise tests of the performance of different
jet algorithms were performed. New precise jet measurements were
presented which will help to constrain further the proton PDFs when
included in global fits.

\end{document}